\documentclass{aa}

\usepackage{psfig}

\def\src  {SGR~1806--20~}
\def\integral {{\it INTEGRAL}~}

\begin{document}



\title{Two years of INTEGRAL monitoring of the Soft Gamma-Ray Repeater SGR 1806-20: from quiescence to
frenzy \thanks{Based on observations with INTEGRAL, an ESA project with instruments and science data centre funded by ESA member states (especially the PI countries: Denmark, France, Germany, Italy, Switzerland, Spain), Czech Republic and Poland, and with the participation of Russia and the USA.} }

   \author{D. G\"{o}tz\inst{1}, S. Mereghetti\inst{1}, S. Molkov\inst{2,3},
    K. Hurley\inst{4}, I.~F. Mirabel\inst{5}, R. Sunyaev\inst{2,3}, G. Weidenspointner\inst{6},
    S. Brandt\inst{7}, M. Del Santo\inst{8}, M. Feroci\inst{8}, E. G{\" o}{\u g}{\" u}{\c s}\inst{9}, A. von Kienlin\inst{3}, M. van der Klis\inst{10}, C. Kouveliotou\inst{11,12}, N. Lund\inst{7}, G. Pizzichini\inst{13}, P. Ubertini\inst{8}, C. Winkler\inst{14}, \and P.~M. Woods\inst{15}
    }

   \offprints{D. G\"{o}tz, email: diego@mi.iasf.cnr.it}

   \institute{INAF -- Istituto di Astrofisica Spaziale e Fisica Cosmica,
              Sezione di Milano ``G.Occhialini'',
          Via Bassini 15, I-20133 Milano, Italy
          \and
      Space Research Institute, Russian Academy of Sciences, Profsoyuznaya 84/32, 117997 Moscow, Russia
      \and
      Max Planck Institut f\"ur Astrophysik, Karl Schwarzschild Str. 1, 85740 Garching bei M\"unchen, Germany
      \and
         University of California at Berkeley, Space Sciences Laboratory, Berkeley CA 94720-7450, USA
     \and
     European Southern Observatory, Alonso de Cordova 3107, Santiago, Chile
     \and
     External ESA Fellow - Centre d'\'Etude Spatiale des Rayonnements (CESR), avenue du Colonel-Roche 9, F-31028 Toulouse Cedex 4, France
     \and
     Danish National Space Center (DNSC), Juliane Maries Vej 30, 2100 Copenhagen, Danemark
     \and
     INAF -- Istituto di Astrofisica Spaziale e Fisica Cosmica, Sezione di Roma, Via Fosso del Cavaliere 100, I-00133 Roma, Italy
     \and
     Sabanci University, Orhanli-Tuzla, Istanbul, Turkey
     \and
     Astronomical Institute "Anton Pannekoek", University of Amsterdam and Center for High-Energy Astrophysics, Kruislaan 403, NL 1098 SJ Amsterdam, The Netherlands
     \and
     National Space Science and Technology Center, 320 Sparkman Drive, Huntsville, AL 35805
     \and
     NASA Marshall Space Flight Center, XD 12, Huntsville, AL 35812
     \and
     INAF -- Istituto di Astrofisica Spaziale e Fisica Cosmica, Sezione di Bologna, Via Gobetti 101, I-40129 Bologna, Italy
     \and
     ESA-ESTEC, RSSD, Keplerlaan 1, 2201 AZ Nordwijk, The Netherlands
     \and
     Department of Physics, University of Manchester, Sackville Street, PO Box 88, Manchester M60 1QD, UK
     }


\abstract{
\src has been observed for more than 2 years with the \integral satellite. In this period the source went
from a quiescent state into a very active one culminating in a giant flare on December 27 2004. Here
we report on the properties of all the short bursts detected with \integral before the giant flare.
We derive their number-intensity distribution and confirm the hardness-intensity correlation for the
bursts found by \cite{gotz04}.
Our sample includes a very bright outburst that occurred on October 5 2004, during which over one hundred bursts were emitted in 10 minutes, involving an energy release of 3$\times$10$^{42}$ erg. We present a detailed analysis of it and discuss our results in the framework of the magnetar model.
}

\authorrunning{D. G\"{o}tz et al.}
\titlerunning{SGR 1806-20 bursts observed with \integral}
\maketitle

\section{Introduction}

SGR 1806--20 is currently one of the most active Soft Gamma-ray Repeaters.
These sources (see \cite{hurleysgrrew,woodsrew} for recent reviews)
were discovered by their recurrent
emission of soft ($\leq$100 keV) gamma-ray bursts.
They undergo sporadic, unpredictable periods of activity,
which last days to months, often followed by long periods
(up to years or decades) during which no bursts are detected.
These recurrent bursts have typical durations of $\sim$0.1 s
and luminosities in the range 10$^{39}$-10$^{42}$ erg s$^{-1}$.
Occasionally, SGRs also emit giant flares that last up to a
few hundred seconds and exhibit remarkable pulsations that
reveal their spin periods (e.g. \cite{mazets}, \cite{hurley1999}, \cite{rhessigiant}).

The spin period of SGRs can also be measured in their persistent (quiescent)
X-ray emission. Typical luminosities  (0.5-10 keV)
of these heavily absorbed sources are of the order of a few 10$^{35}$ erg s$^{-1}$.
Persistent emission from \src has been detected recently also in the soft
$\gamma$-ray range up to $\sim$150 keV (\cite{mereghetti05,molkov}).

The bursting activity and the persistent emission
are generally explained
in the framework of the ``magnetar'' model (see e.g. \cite{dt92}, \cite{pac92},
\cite{td95}), which involves slowly
rotating ($P\sim$ 5-8 s) highly magnetised ($B\sim$10$^{15}$ G) isolated neutron stars.
The magnetar model is based on the fact that the rotational energy loss inferred
from the spindown (in the 10$^{-10}$--10$^{-13}$ s s$^{-1}$ range)
is not sufficient to power the persistent X-ray luminosity. Hence
it is the decay of the magnetic field itself that provides the necessary energy. In this
framework the magnetic dissipation causes the neutron star crust to fracture.
These fractures generate sudden shifts in the magnetospheric footprints, which
trigger the generation of Alfv\'{e}n waves, which in turn accelerate
electrons above the pair-production threshold, resulting quickly in
an optically thick pair-photon plasma. The cooling of this plasma
generates the typical short bursts of soft $\gamma$-ray radiation.
An alternative explanation for short bursts, proposed by \cite{lyutikov},
does not involve crustal fractures but simply local magnetic reconnection.
The longer, much more energetic and rare
flares are powered by a sudden reconfiguration of the magnetic field
through magnetic reconnection (as in solar flares), and involve the entire neutron
star magnetosphere.

After a period of quiescence, \src became active in the summer of 2003  (\cite{hurley2003}).
Its activity then increased in 2004 (see e.g. \cite{moresgr,evenmoresgr}).
A strong outburst
(\cite{bigone1}) during which about one hundred short bursts were emitted in a few minutes
occurred on October 5 2004 (see Section \ref{thebigone}).
Finally a giant flare,
whose energy ($\sim$10$^{46}$ erg) was two orders of magnitude larger than those of the
previously recorded flares from SGR 0526-66 and SGR 1900+14,
was emitted on December 27$^{th}$ 2004 (see e.g. \cite{swiftgiant,acsgiant,rhessigiant}).

Here we report on all the short bursts detected by \integral in
2003 and 2004.

\section{Observations and data analysis}

\begin{figure}[ht!]
\centerline{\psfig{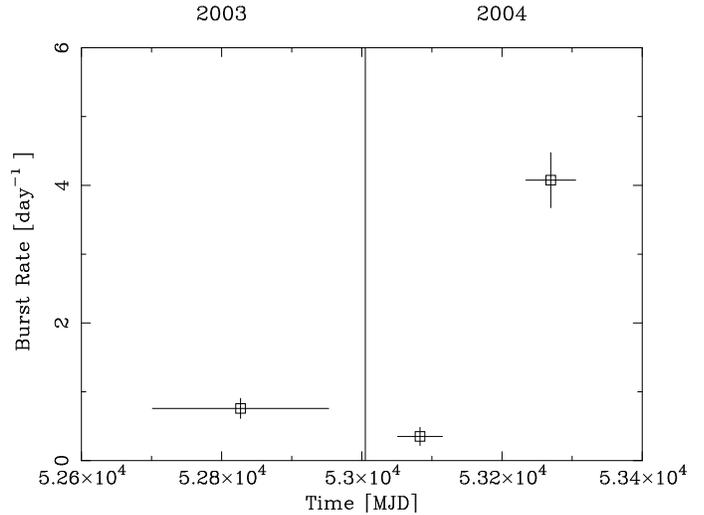}}
\caption{\src burst rate vs. time.  Only the bursts with
fluence $\geq$ 3$\times$10$^{-8}$ erg cm$^{-2}$ are considered.
The three time periods correspond roughly to Fall 2003,
Spring 2004 and Fall 2004.}
\label{sgrrate}
\end{figure}

The observations of this source obtained with the
\integral satellite in September and October 2003,
during a period of moderate bursting activity,
have been reported earlier (see \cite{gotz04,gotzmunich}).
We have re-analysed those data and, in addition, we report here
on a much larger burst dataset.
Excluding the bursts of the October 5$^{th}$ event (see below),
we have detected and analysed 224 bursts.
The observations are summarised in Tab. \ref{sgrobs}.
\begin{table*}[ht]
\begin{center}
\begin{tabular}{|c|c|c|c|c|}
\hline
INTEGRAL & Obs.  & Number    & Exposure   & Orbit Start\\
Orbit    & Type & of Bursts & Time [ks] &Time [UT]\\
\hline
56 & P & 1 &  90.1  & 2003-03-29T21:37:59\\
105 & P & 8 & 207.8 & 2003-08-23T10:24:56 \\
108 & P & 2 & 215.5 & 2003-09-01T09:41:20\\
109 & P & 2 & 217.2 & 2003-09-04T09:26:46\\
114 & P & 1 & 208.7 & 2003-09-19T08:27:22\\
119 & P & 1 & 74.8  & 2003-10-04T07:22:32\\
120 & P & 4 & 201.2 & 2003-10-07T07:11:55\\
121 & P & 2 & 140.7 & 2003-10-10T07:02:43\\
122 & P & 22& 201.3 & 2003-10-13T06:53:37\\
171 & CP & 1& 197.9 & 2004-03-07T22:21:10\\
173 & CP & 2 & 175.5& 2004-03-13T21:54:43\\
175 & CP & 2 & 72.1 & 2004-03-19T21:30:35\\
181 & CP/R & 3 & 134.8 & 2004-04-06T20:15:13 \\
\hline
56--181&&51&2138&\\
\hline\hline
225 & TOO & 41 & 169.9 &2004-08-16T09:51:26\\
226 & CP & 5    & 32.1  &2004-08-19T09:37:14\\
227 & R & 59    & 148.2 &2004-08-22T09:22:35\\
229 & CP & 1    & 68.1  &2004-08-28T08:53:12 \\
230 & R & 2     & 208.5 &2004-08-31T08:41:37\\
232 & CP & 1    & 52.1  &2004-09-06T08:17:45\\
233 & CP & 2    & 36.2  &2004-09-09T08:04:02\\
234 & CP & 1    & 14.8  &2004-09-12T07:50:28 \\
235 & CP & 2    & 77.9  &2004-09-15T07:37:33\\
236 & R & 8     & 178.1 &2004-09-18T07:24:13\\
237 & R & 2     & 173.2 &2004-09-21T07:09:49\\
240 & CP & 8    & 208.7 &2004-09-30T06:34:39\\
241 & CP & 9$^{1}$ & 177.3&2004-10-03T06:23:00\\
242 & CP & 2    & 89.3 &2004-10-06T06:10:07\\
243 & CP & 4    & 107.0 &2004-10-09T05:56:48\\
244 & CP & 1    & 69.0 &2004-10-12T05:43:59\\
245 & R & 9     & 157.3 & 2004-10-15T05:31:01\\
246 & R & 14    & 108.0 &2004-10-18T05:16:53\\
249 & CP & 2    & 111.1& 2004-10-27T04:42:52\\
\hline
225--249&&173&2177&\\
\hline
\end{tabular}
\end{center}
$^{1}$ Excluding the huge outburst of October 5 (see text)
\caption{Observation summary of the bursts from \src analysed here.
(P: Public Data, CP: Core Program Observations, TOO: Target of Opportunity Observations,
R: Open time Galactic Centre Data)
The double horizontal line indicates the separation between the periods of different
activity of the source (see text).}
\label{sgrobs}
\end{table*}
Our dataset represents the largest sample of short bursts imaged from \src
to date.
As can be seen from Fig. \ref{sgrrate}, the burst activity of \src increased significantly
with time, and the source was particularly active during Fall 2004.



Here we concentrate on the data of the \integral imager IBIS (\cite{ibis}).
We do not use the data of the X-ray monitor JEM-X (\cite{jemx}), since
only a small fraction of the bursts were inside its smaller field of
view and, owing to their brightness, they were affected by saturation and dead time
effects. All the bursts were in the field
of view of the SPI instrument (\cite{spi}), but at energies below $\sim$300 keV SPI
has a much smaller effective
area compared to IBIS, which is therefore the most suitable instrument for detailed studies
of short bursts.

IBIS is a coded mask telescope with a large field of view
(29$^{\circ}\times$29$^{\circ}$ zero sensitivity, 9$^{\circ}\times$9$^{\circ}$
full sensitivity).
We used the low energy (15 keV-1 MeV) detector layer ISGRI (\cite{isgri}),
composed of 128$\times$128 independent CdTe detectors (pixels)
yielding a geometric area of $\sim$1300 cm$^{2}$ on axis.
ISGRI operates
in a {\it photon-by-photon} mode: this means that for each photon the energy, arrival
time and interaction pixel are known.
To identify the bursts we used the trigger information provided by the \integral Burst Alert
System (IBAS; \cite{ibas}) for the Target of Opportunity (ToO), public and Core Program data.
For the Galactic Centre Field data, we searched for the bursts by computing
light curves with 10 ms time resolution and looking for significant excesses
corresponding to the direction of \src (for details see \cite{molkov}).
All the bursts are unambiguously associated, within
the typical $\sim$2$^{\prime}$ location uncertainties, with the position
of the X-ray counterpart of \src (\cite{chandra}).

For each burst we derived the 15-100 keV light curve with 10 ms time resolution.
This energy band was chosen because no emission above 100 keV is
detected in the typical short bursts.
In order to increase the signal-to-noise
ratio, the light curves have been extracted by selecting
only the pixels illuminated by the source over at least half their surface.
The background was estimated by fitting a constant
to 2 s time intervals before and after the burst.
The light curves were then corrected for vignetting, caused by
the fact that the bursts were detected at different off-axis angles, and
for ISGRI dead time. The detector
is made of 8 identical modules and the good events dead time, $\tau$, for each module
is $\sim$114 $\mu$s. For each burst, based on its position in the IBIS field of view,
we determined the number of modules involved in the detection for at least half their surface, $M$,
and  applied the following dead time correction

\begin{equation}
N_{inc}=\left(\frac{1}{N_{rec}}-\frac{\tau}{M}\right)^{-1},
\end{equation}

where $N_{rec}$ is the recorded flux (cts/s) in each time bin and N$_{inc}$ is the reconstructed
incident flux. By evaluating this relation, one sees that only for the bins with a ratio between
the recorded flux and the maximum recordable flux ($M/\tau$) larger than 0.3 a significant
correction is required. For the rest of the bins the dead time effect is less than a few percent.
It turns out that only 5.3\% of the bursts have peak fluxes above this limit and
their light curves have been corrected accordingly.

These light curves have been used to determine
the $T_{90}$ durations (i.e. the time during which 90\% of the burst fluence
is accumulated), the peak fluxes (counts/bin) and the fluences (counts) of each burst.
To convert the peak fluxes and fluences to physical units we have used the conversion factor
1 count s$^{-1}$ = 1.5$\times$10$^{-10}$ erg cm$^{-2}$ s$^{-1}$,
derived by the spectral analysis of the brightest bursts (see \cite{gotz04}).
These bursts are well fit with a
Thermal Bremsstrahlung (TB) model with temperatures (kT) between 30 and 40 keV.
The above conversion factor is valid for the average temperature of 32 keV. A
variation of the temperature between the two extremes implies a variation of only 7\%
of the conversion factor. So we have applied it also to fainter bursts, under
the assumption that the spectral shape is the same for all
the bursts. We point out, however, that extracting the average spectrum of
the 2003 bursts and comparing it to the average spectrum of a representative
subsample of the bursts of Fall 2004, during which the source was more active,
we find that the latter is marginally harder, but still within the above mentioned limits.
Both average spectra are well represented by the TB model.

\section{Results}

\subsection{Spectral Evolution}

Evidence for spectral evolution within the bursts of \src
was reported for the first time by \cite{gotz04}, who found that
some bursts show significant spectral evolution, while
others, particularly those with a  ``flat topped'' profile, do not.
In addition, a hardness-intensity
anti-correlation within the bursts of \src was reported by the
same authors.

\begin{figure}[ht!]
\centerline{\psfig{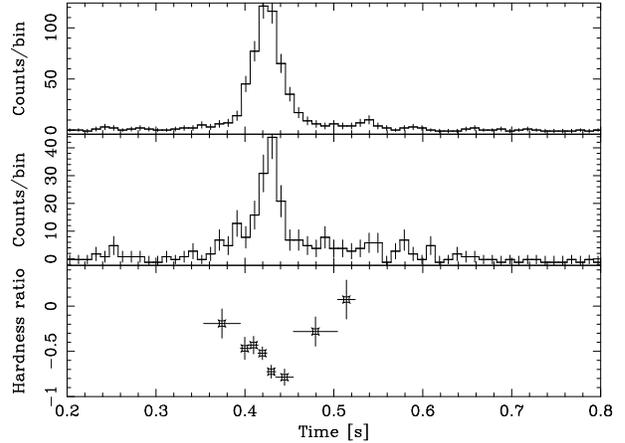}}
\caption{IBIS/ISGRI light curves in the soft (20-40 keV, upper
panel) and hard (40-100 keV, middle panel) energy range and
hardness ratio (lower panel) for a short burst from SGR 1806--20.
Time=0 corresponds to 2004-08-18T21:52:49.39 UT}
\label{ev1}
\end{figure}

\begin{figure}[ht!]
\centerline{\psfig{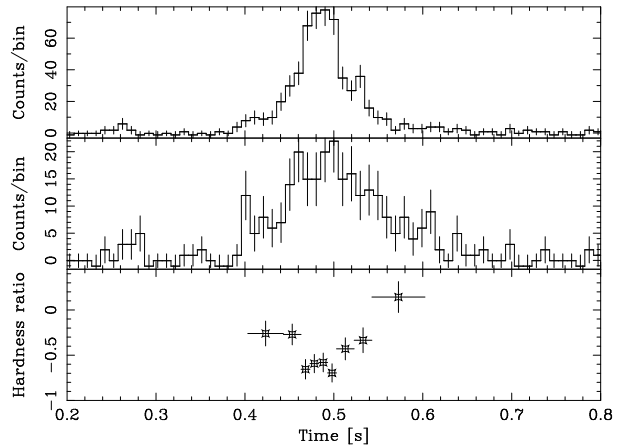}}
\caption{IBIS/ISGRI light curves in the soft (20-40 keV, upper
panel) and hard (40-100 keV, middle panel) energy range and
hardness ratio (lower panel) for a short burst from SGR 1806--20.
Time=0 corresponds to 2004-08-21T03:38:26.31 UT.}
\label{ev2}
\end{figure}

\begin{figure}[ht!]
\centerline{\psfig{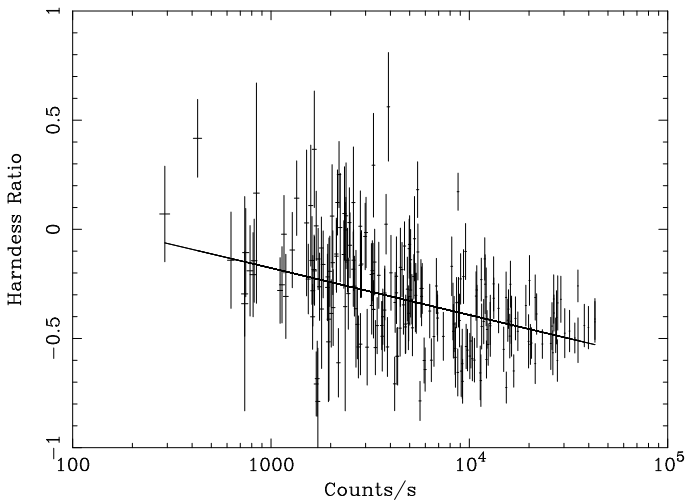}}
\caption{Hardness ratio (($H$--$S$)/($H$+$S$)) versus total count rate (20-100 keV,
corrected for vignetting). The points are derived from the time
resolved hardness ratios of the bursts with the best statistics.
The line indicates the best fit with a linear function given in
the text.}
\label{hi}
\end{figure}

We investigated the spectral evolution of the
brightest bursts by computing the background subtracted light
curves in two different energy bands
(20-40 ($S$) and 40-100 ($H$) keV)
and evaluating the hardness ratio, $HR$=($H$--$S$)/($H$+$S$), in time bins with the same number of total
counts. The energy bands used here are slightly different from those used in \cite{gotz04}
because we realized that the individual pixels behave differently at energies between
15 and 20 keV.
In our analysis the spectral evolution is confirmed: excluding ``flat topped bursts",
peaks tend to be softer than tails. Two good examples are shown
in Figs. \ref{ev1} and \ref{ev2}.
However, a number of bursts
show a less certain spectral evolution. By considering
all the individual bins of all the bursts we derived
the hardness ($HR$) intensity ($I$) distribution shown in Fig. \ref{hi}.
To investigate if the data are correlated, we computed the
Spearman rank-order correlation coefficient of the 217 data points, $R_{s}$,
which is --0.49. This corresponds to a chance probability of
4$\times$10$^{-15}$ (7.4 $\sigma$) that our distribution is due to uncorrelated data.
According to an F-test the data are significantly (8$\sigma$) better described by a
linear fit ($HR=0.47 - 0.22\times\log(I)$) than by a constant. The parameters of the
fit are very similar to those found by \cite{gotz04}.
So we can conclude that an anti-correlation between hardness and
intensity is confirmed by our data. The inclusion or exclusion of flat-topped events
from our analysis does not affect our results, since they represent a small fraction
of the bursts.

\subsection{The Large Outburst of October 5 2004}

\label{thebigone}
On October 5 2004 IBAS triggered at 13:56:49 UT on the first
of a series of bursts emitted from \src. The  source remained
active, emitting several tens of bursts, until 14:08:03 UT.

\begin{figure*}[ht!]
\centerline{\psfig{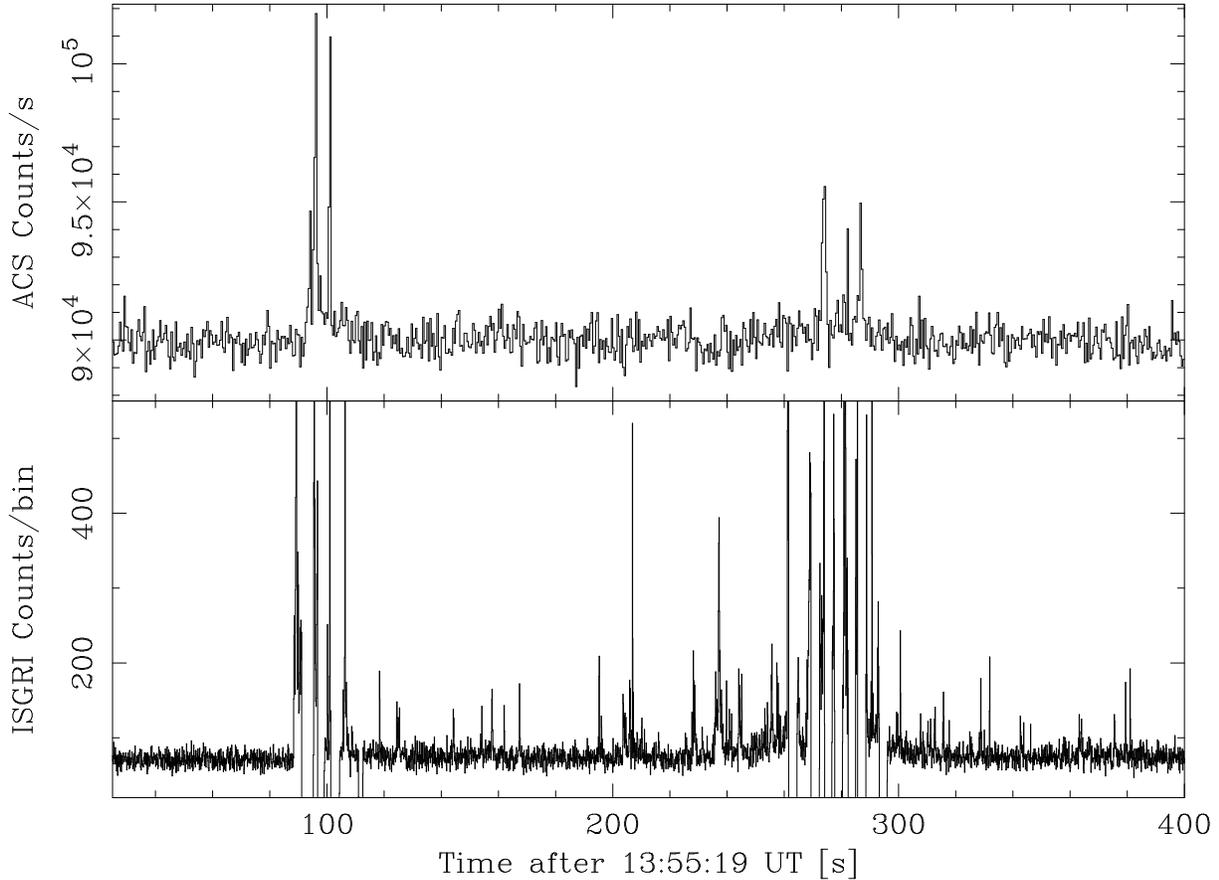}}
\caption{Light curves of the initial part of the October 5, 2004 outburst
 of SGR 1806-20. Upper panel: light curve at energy
 greater than $\sim$80 keV obtained with the SPI Anti-Coincidence System in bins of 0.5 s.
Bottom panel: light curve in the 15-200 keV energy range obtained with
  the IBIS/ISGRI instrument (bin size 0.1 s). The gaps in the IBIS/ISGRI light
  curve are due to saturation of the satellite telemetry.}
\label{biglc}
\end{figure*}

\begin{figure*}[ht!]
\centerline{\psfig{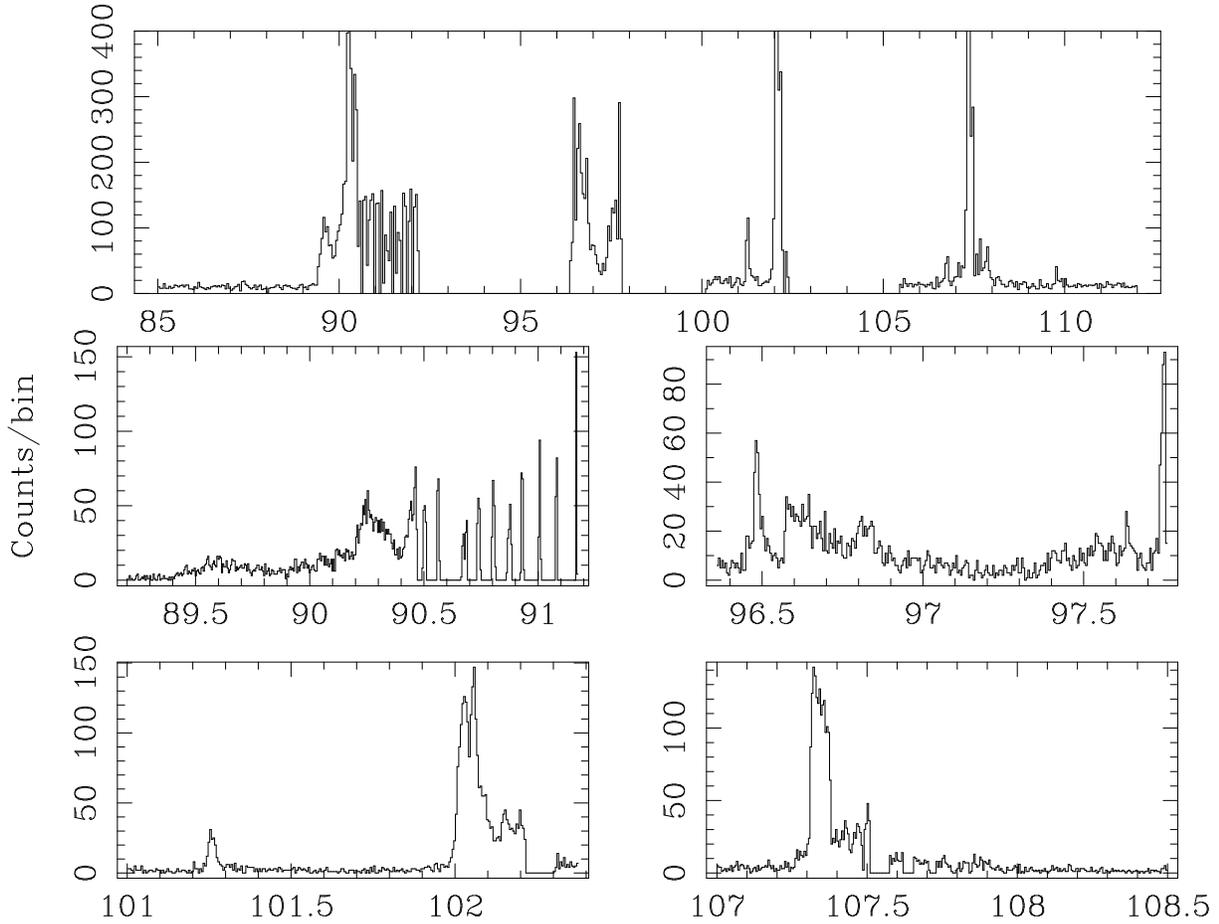}}
\caption{IBIS/ISGRI light curves (15-200 keV) of the first cluster of bright bursts. Upper Panel: overall
light curve with 0.05 s resolution. The lower panels show some parts of the light curve in detail
with a time resolution of 5 ms. The time scale is in seconds and is the same as for Fig. \ref{biglc}. }
\label{zoom1}
\end{figure*}

\begin{figure*}[ht!]
\centerline{\psfig{figure=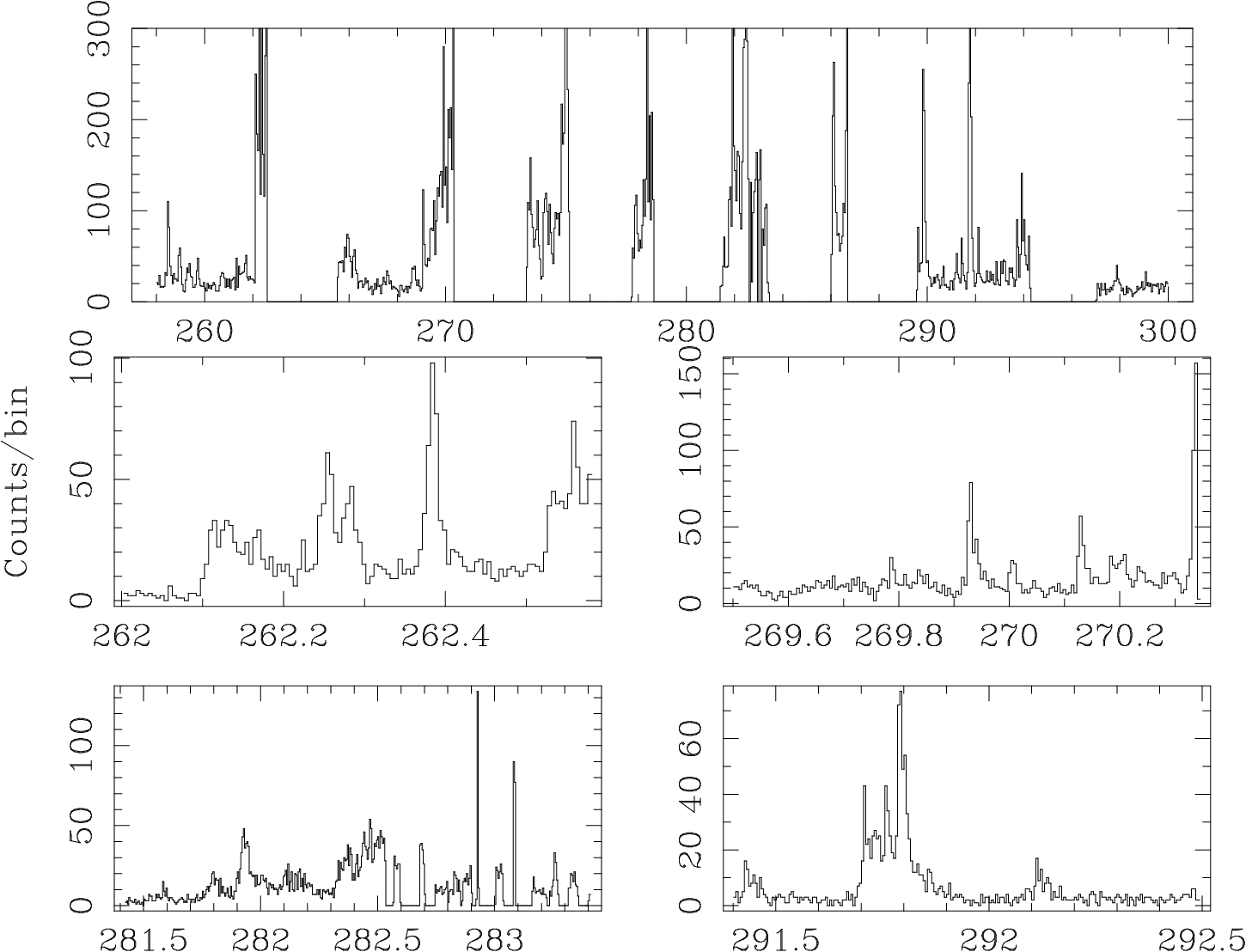,width=16cm,angle=-90}}
\caption{IBIS/ISGRI light curves (15-200 keV) of the first cluster of bright bursts. Upper Panel: overall
light curve with 0.05 s resolution. The lower panels show some parts of the light curve in detail
with a time resolution of 5 ms. The time scale is in seconds and is the same as for Fig. \ref{biglc}.}
\label{zoom2}
\end{figure*}

Fig. \ref{biglc} (lower panel) displays the initial and
brighter part of the $\sim$ 11 minute long outburst. The time scale
starts at 13:55:19 UT. We will refer to this time scale throughout our analysis.
Two bright clusters of bursts are visible at t$\sim$100 s and t$\sim$280 s.
They are so bright that the satellite telemetry was partially saturated
(0 counts in the plot) for about 10 and 20 seconds respectively
and only part of the registered events could be sent to the ground.
The two bright clusters are shown in more detail in Figs. \ref{zoom1} and \ref{zoom2}.
Despite the presence of the telemetry gaps in the data, one can see that they consist of many
short bursts, with significant variability down to $\sim$10 ms time scales.
At least 75 short bursts, typical in terms of duration and shape, have been
detected in IBIS/ISGRI data. 

\begin{figure}[ht!]
\centerline{\psfig{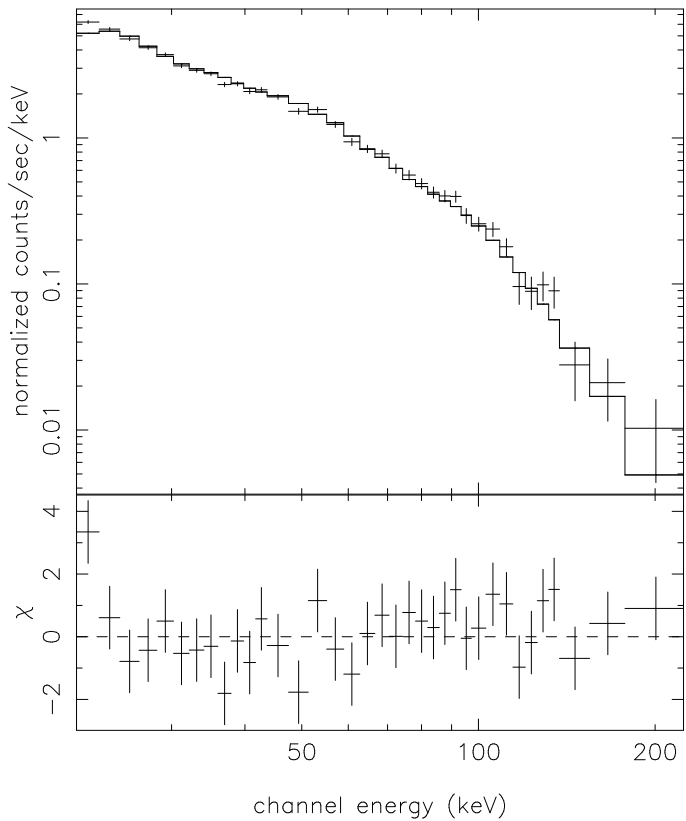}}
\caption{Spectrum of the October 5 outburst of SGR 1806--20
 measured with the IBIS/ISGRI instrument. Upper panel: data points and
 best fit with a thermal bremsstrahlung model with kT=58 keV.
 Lower panel: residuals from the best fit in units of sigma.}
\label{spectrum}
\end{figure}

We have extracted a spectrum over the entire duration of the outburst
(676 s integration time, see Fig. \ref{spectrum}).
A fit in the 20-300 keV band with a thermal bremsstrahlung model
using XSPEC v11.2 (\cite{xspec})
gives a good result ($\chi^{2}$ = 39.64 for 40 dof, allowing for 5\% systematics) and
yields a temperature of 58$\pm$2 keV, higher
than that usually  measured for short bursts. A slightly worse but still
acceptable fit ($\chi^{2}$ = 45.99 for 40 dof) can be obtained
using the sum of two black
bodies, as proposed by \cite{feroci} for the cumulative spectrum
of short bursts from SGR 1900+14 and by \cite{olive} for an intermediate duration
burst from the same source. The temperatures of the two black bodies
are kT$_{1}$=5.8$\pm$0.4 keV and kT$_{2}$=18$\pm$0.9 keV.
Both values are higher than those
derived by \cite{feroci} (kT$_{1}$=3.3$\pm$0.1 keV,
kT$_{2}$=9.5$\pm$0.4 keV) and by \cite{olive} (kT$_{1}$=4.3$\pm$0.1 keV,
kT$_{2}$=9.9$\pm$0.3 keV). On the other hand by fixing the temperatures to the
values derived by these authors we get unacceptable fits.
This is possibly due to the fact that their
spectra extend to lower energy. In the ISGRI energy range
this model 
does not improve the reduced $\chi^{2}$ of the fit 
and hence the results
of the two black body fit have to be taken with care.


We computed the hardness ratios over the time periods of the two clusters of bright
bursts (t = 88--108 s and t = 260--293 s
respectively). They indicate
a clear hardening for the second cluster with respect to the first one.
Even though the ISGRI detector partially saturates over these
time intervals, the spectra still have enough counts to perform spectral analysis
(see Figs. \ref{zoom1} and \ref{zoom2}).
We have hence extracted two spectra over these two time periods.
Applying TB models to the spectra, we see that the temperature increases significantly
with time. In fact for the first cluster of bursts we derive a temperature (kT) of 39$\pm$2 keV, while
for the second cluster we get 55.5$\pm$1.5 keV. We point out however (see Tab. \ref{temptab})
that the TB model does not fit the spectrum of the first cluster well. We divided this
period into smaller time intervals and found that it is only the initial part (t = 89--92 s),
that cannot be represented by a TB model. Its spectrum
is in fact much better fitted ($\chi^{2}$/dof = 36.6/31) by a power law with
photon index $\Gamma$=2.80$\pm$0.05. The overall spectrum is hence a mixture of different
states, which may also explain the failure of the two black body model.

We have extracted 5 more spectra
for 2 bright bursts at  t = 205.45 and t = 237 s 
and for 3 short ($\sim$0.1 s) bursts at t = 406.4, 560.25 and 761.8 s respectively.
All the spectra are better represented by TB models than by power laws. The derived temperatures
are reported in Tab. \ref{temptab}. Although no specific trend can be
identified, there are significant variations
between the bursts' temperatures. The spectra of the bursts at t = 205.45, 237 and 560.25 s
have very high temperatures, above
40 keV, which is the maximum temperature usually measured for short bursts.
These events are similar in spectral hardness to two bursts
detected from SGR 1900+14 following its giant flare (\cite{woods}).

\begin{table}[ht]
\begin{center}
\begin{tabular}{|c|c|c|c|c|}
\hline
Time [s] & $\Delta$t [s] &kT [keV]  & Flux 20-200 keV  & $\chi^{2}$/dof \\
         &              &          & [erg cm$^{-2}$ s$^{-1}$]& \\
\hline
88 & 20 &39$\pm$2&      $>^{1}$1.5$\times$10$^{-7}$  &118/36\\
205.45& 4.1 &61$\pm$8&  9.0$\times$10$^{-8}$ &33/28\\
237 & 4 &74$\pm$7&      1.5$\times$10$^{-7}$ &39/33\\
260& 33 &55.5$\pm$1.5&  $>^{1}$1.5$\times$10$^{-7}$ &52/40\\
406.4&0.2&34.5$\pm$11&  1.5$\times$10$^{-7}$&6.8/6\\
560.25&0.45&65$\pm$7&   2.6$\times$10$^{-7}$ &23/20\\
761.8&0.2&44$\pm$7&     2.8$\times$10$^{-7}$  &19/16\\
\hline
\end{tabular}
$^{1}$The fluxes of the two bright clusters are not corrected for
telemetry saturation.
\end{center}
\caption{Temperatures, fluxes and statistical goodness of the fits of the different
bursts obtained with the TB model. }
\label{temptab}
\end{table}

The fluence of the entire outburst as measured by ISGRI is 1.5$\times$10$^{-5}$ erg cm$^{-2}$. This
value is however heavily affected by the saturation of the brightest bursts and represents only
a lower limit to the real fluence.
In order to derive the overall fluence, we used the data provided
by the Anti-Coincidence System (ACS) of the \integral spectrometer (SPI).
As can be seen in Fig. \ref{biglc} (upper panel),
only the brightest bursts are visible in these data and hence they represent
complementary information to the ISGRI data.

ACS data consist of the total count
rate (50 ms time resolution) above 80 keV measured by the 91 bismuth germanate (BGO)
scintillator crystals  that surround the \integral spectrometer.
The crystals are used as the anti-coincidence system of the spectrometer but are also capable
of detecting high-energy transient events such as bright GRB and SGR bursts.
We used the Monte Carlo package MGGPOD (\cite{weid05}) and a detailed mass modelling of SPI and the whole
satellite (\cite{weidenspointner} and references therein) to derive the effective area of the ACS
for the direction of \src.
We computed the ACS light curve with a binsize of 0.5 s and estimated the background by fitting a constant value to all the data of the same pointing excluding the bursts.
We used the background subtracted light curve to compute the fluence of each burst cluster
in counts. The ACS data do not provide any spectral information, so we computed
the conversion factor to physical units based on the spectral
shapes derived from ISGRI data and on the effective
area computed through our simulations.
The resulting fluences
above 80 keV are 1.2$\times$10$^{-5}$ and 9.4$\times$10$^{-6}$ erg cm$^{-2}$ for the first
and second clusters respectively. Converting these fluences to the 15-100 keV band
one obtains 7.4$\times$10$^{-5}$ and 3.2$\times$10$^{-5}$ erg cm$^{-2}$ respectively. By adding these results to the ones obtained for the
ISGRI total spectrum, one can derive the total energy output during the whole event, which is
1.2$\times$10$^{-4}$ erg cm$^{-2}$. This corresponds to 3.25$\times$10$^{42}$ erg
for an assumed distance of 15 kpc (\cite{corbel}).

\subsection{Number-Intensity Distribution}

The fluences of the 224 bursts derived above (excluding the ones
detected during the October 5$^{th}$ event) have been used to compute the
number-intensity distribution (Log N-Log S) of the bursts.

\begin{figure}[ht!]
\centerline{\psfig{figure=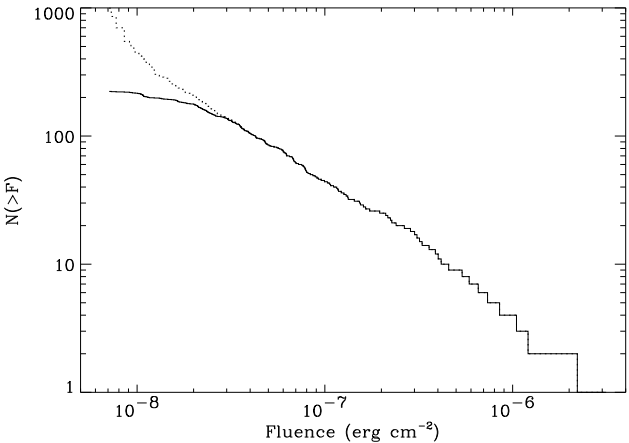,width=9cm}}
\caption{Number-intensity distribution of all the bursts detected by \integral in 2003 and 2004.
The continuous line represents
the experimental data, while the dashed line represents the data corrected
for the exposure.}
\label{logn}
\end{figure}

The experimental distribution deviates significantly from
a single power-law (Fig. \ref{logn}). This is first of all due to the fact that the source has
been observed at different off-axis angles.
The faintest bursts are missed when the source is observed at large off-axis angles.
In order to correct for this effect we have computed the effective exposure of the source,
taking into account the variation of sensitivity at various off-axis angles.
This yields the exposure-corrected cumulative distribution shown by
the dashed line in Fig. \ref{logn}.


Since the numbers at each flux level are not statistically independent, one cannot use
a simple $\chi^{2}$ minimisation approach to fit the cumulative number-intensity distribution.
So we have used the unbinned detections and applied the Maximum Likelihood
method (\cite{crawford}), assuming a single power-law distribution for the number-flux
relation ($N(>S)\propto S^{-\alpha}$). We have
used only the part of the distribution
where completeness was achieved (i.e. $S\geq$3$\times$10$^{-8}$ erg cm$^{-2}$).
In this case the expression to be maximised is
\begin{equation}
{\cal L} = T \ln \alpha - \alpha\sum_{i}\ln S_{i}-T\ln (1-b^{-\alpha})
\end{equation}
where $S_{i}$ are the unbinned fluxes,
$b$ is the ratio between the maximum and minimum values of the
fluxes, and $T$ is the total number of bursts.
This method yields $\alpha$=0.91$\pm$0.09.
If a single power-law model is an adequate representation of data,
the distribution of the quantities
\begin{equation}
y_{i}=\frac{1-S_{i}^{-\alpha}}{1-b^{-\alpha}}
\end{equation}
should be uniform over the range (0,1). In our case, a Kolmogorov-Smirnov (K-S) test
shows that a power law is an appropriate model, yielding a probability of 98.8\% that
the data are well described by our model.

We then divided the bursts in two samples comprising 51 and 173 bursts respectively.
The division is based on the periods of different activity.
The first period ends at orbit 181 and the second one starts at
orbit 225 (see Tab. \ref{sgrobs}). The two slopes derived with the Maximum Likelihood method
are  $\alpha$=0.9$\pm$0.2 for the low level activity period and
$\alpha$=0.88$\pm$0.11 for the high level one. The two slopes are
statistically consistent with each other and a K-S
test shows that the probability that the two distributions are drawn
from the same parent distribution is 93\%. Thus we conclude that the
the relative fraction of bright and faint bursts is not influenced
by the level of activity of the source.

\section{Discussion}

The bursts detected by \integral and presented here are common bursts as
far as many aspects are concerned, such as
durations ($\sim$0.12 s on average), peak fluxes, fluences and spectra
but are the faintest detected at these energies. Thanks to imaging we are
confident that all of them were emitted by \src.

The good quality of our data has allowed us to study these bursts in detail.
In particular, with respect to previous experiments, we have a better combination of
sensitivity, timing and spectral capabilities also for faint events.
Hence we have been able to confirm the early findings
(see \cite{gotz04}) on the spectral evolution of \src short bursts. The fact that burst peaks tend to
be softer than their tails leads to an overall hardness-intensity anti-correlation.
This characteristic still has no clear explanation within the magnetar model.

The relatively large number of bursts has allowed us to constrain
the shape and slope of the fluence distribution, which is well
described by a single power-law on  over 2.5 orders of magnitudes.
The INTEGRAL LogN-LogS has a slope of 0.91$\pm$0.01, similar to
that derived at larger fluences (10$^{-7}$--10$^{-5}$ erg
cm$^{-2}$) with {\it KONUS} data (\cite{konus}). This slope is
also compatible, within 2 $\sigma$,  with the 0.71$\pm$0.11
obtained by combining BATSE and {\it ICE} data (\cite{gogus2000})
over the 5$\times$10$^{-8}$--10$^{-5}$ erg cm$^{-2}$ fluence
range. On the other hand our LogN-LogS is significantly steeper
than that obtained with {\it RXTE}, which has a slope of
0.43$\pm$0.06 over the 10$^{-10}$--10$^{-7}$ erg cm$^{-2}$ fluence
range, (\cite{gogus2000}).
The difference between our slope and the {\it RXTE} one, which
extends to lower fluences, is significant: this may imply that
there is a break in the distribution, but we point out that in the
overlapping part of the two datasets (6$\times$10$^{-9}$ $< S <$
10$^{-7}$ erg cm$^{-2}$) the {\it RXTE} slope is clearly
statistically rejected by our data. We can therefore conclude that,
considering IBIS and BATSE/{\it ICE} data, a single power-law gives
a good representation of the cumulative energy distribution of
\src bursts over 3.5 orders of magnitude.

Persistent emission in the 20-150 keV range has been detected from
\src by integrating over long time intervals part of the
IBIS/ISGRI data reported here (\cite{mereghetti05,molkov}).
Although the persistent emission has a spectrum harder than that
of typical bursts, it is worth to check  whether a significant
part of such persistent emission, could  be due to the cumulative
contribution of numerous bursts too faint to be detected
individually. Integrating our LogN-LogS distribution down to
S$\sim$10$^{-10}$ erg cm$^{-2}$ we find that the burst contribution is at most a few
percent ($\sim$10$^{-12}$ erg cm$^{-2}$ s$^{-1}$) of the total.


The hard and very energetic outburst of October 5$^{th}$ resembled
a similar event seen in SGR 1900+14 by {\it KONUS} on May 30 1998
(\cite{konus}). In that case the outburst lasted 5 minutes and a
total of 26 bursts were detected. The fluence was
3.5$\times$10$^{-4}$ erg cm$^{-2}$, similar to what we measure
here, and the temperatures of the single bursts showed no
particular correlation with time, but were on average at the high
end of their distribution. In both sources, these energetic
outbursts formed by a rapid sequence of relatively hard short
bursts, preceded of a few months the occurrence of giant flares.
The October 5$^{th}$ event fits in the trend of increasing source
activity shown by \src in the last two years and also manifested
in the rise in luminosity  and spectral hardness of the persistent
emission at high (20-150 keV, \cite{mereghetti05}) and low (2-10
keV, \cite{xmm}) energies. On the other hand, this peculiar event
did not mark a  peak or a turnover in the SGR activity. In fact
the two {\it XMM} observations of  \src performed just before
(September 6 2004) and the day after this large outburst (as a ToO
in response to it) yielded similar spectral parameters, fluxes and
pulse profiles, and bursts were seen in both observations
(\cite{xmm}). Also the INTEGRAL data indicate  that the bursting
activity remained high after this event. This also happened in
1998 to SGR 1900+14, which after the May 30 event remained in a
very active state leading to the August 27 giant flare
(\cite{konus}).

These results can be explained in the framework of a recent
evolution of the magnetar model: \cite{lyutikov} explains SGR
bursts as generated by loss of magnetic equilibrium in the
magnetosphere, in close analogy to solar flares: new
current-carrying magnetic flux tubes rise continuously into the
magnetosphere, driven by the deformations of the neutron star
crust. This in turn generates an increasingly complicated magnetic
field structure, which at some point becomes unstable to resistive
reconnection. During these reconnection events, some of the
magnetic energy carried by the currents associated with the
magnetic flux tubes is dissipated. The large event described here
can be explained by the simultaneous presence of different active
regions (where the flux emergence is especially active) in the
magnetosphere of the neutron star. In fact, a long outburst with
multiple components is explained as the result of numerous
avalanche-type reconnection events, as reconnection at one point
may trigger reconnection at other points. This explains the fact
that the outburst seems to be composed by the sum of several short
bursts. This kind of event might  indicate  a particularly
complicated phase of the magnetic field structure which eventually
led to a global restructuring of the whole magnetosphere with the
emission of the giant flare on December 27. This mode also
suggests that short events are due to reconnection, while longer
events have in addition a large contribution from the surface,
heated by the precipitating particles, and are harder. This may
explain the generally harder spectra observed.

Thus events like these release a small (compared to giant flares)
fraction of the energy stored in the twisted magnetic field of the
neutron star, not allowing the magnetic field to decay
significantly. They are rather related to phases of high activity
due to large crustal deformations (indicating that a large
quantity of energy is still stored in magnetic form) and can be
looked at as precursors of a major reconfiguration of the magnetic
field.


Finally we would like to point out that after the giant flare of
December 27 2004, \src has remained active, with a burst rate per
day of $\sim$ 1.4 between February 16 and April 28 2005. This
preliminary value indicates that the level of activity of the
source is intermediate between 2003 and 2004. The emission of the
giant flare, triggered by the reconfiguration of the magnetic
field, has lowered the crustal stresses due to the magnetic
dissipation (\cite{tlk}). A similar behaviour is seen in the X-ray
band, where the persistent flux is intermediate between the 2003
and the 2004 observations (\cite{xmm,nanda,tiengo}).

\section{Conclusions}

We have presented the results of two years of \integral monitoring
of \src. During this time period the source went from a
$\gamma$-ray quiescent state into a very active one that
culminated in the giant flare of December 27, 2004.

Our good quality data for low fluence bursts allowed us to
establish that the number-density relation of the bursts is well
represented by a single power law with index
$\alpha$=0.91$\pm$0.09. Despite the increase in the rate of
emitted bursts, the burst properties  do not change significantly
with time, neither does the slope of the number-density
distribution. The spectral evolution of the bursts discovered in
the 2003 data is confirmed also by this improved analysis of a
larger sample of 224 bursts. The fact that bursts peaks tend to be
softer than tails results in a hardness-intensity anti-correlation
within the bursts.

On October 5, 2004, close to the peak of its activity \src emitted
within 10 minutes more than a hundred short bursts that had a
spectrum slightly harder than usual and involved a total energy
release of 3$\times$10$^{42}$ erg. These high temperature bursts
can be explained in the framework of the magnetar model as an
avalanche-type reconnection event in the neutron star
magnetosphere caused by a particularly complicated structure of
the magnetic field. Events of this kind might be precursors of
the major reconfiguration of the whole magnetic field of the
neutron star causing large flares in SGRs.

\begin{acknowledgements}
We acknowledge the Italian Space Agency financial and programmatic
support via contract I/R/046/04. SeM thanks RFBR for grant
05-02-17465. KH is grateful for support under the INTEGRAL US
Guest Investigator program, NAG 5 - 13738. The SPI-ACS is
supported by the German "Ministerium f\"{u}r Bildung and
Forschung" through the DLR grant 50.OG.9503.0.
\end{acknowledgements}

\end{document}